\documentclass[reprint,twocolumn,amsmath,amssymb,aps,superscriptaddress]{revtex4-2}

\usepackage{graphicx}
\usepackage{dcolumn}
\usepackage{bm}
\usepackage{graphicx}
\usepackage{amssymb}
\usepackage{epstopdf, epsfig}
\usepackage{yfonts}
\usepackage[colorlinks=true, linkcolor=blue, citecolor=blue, urlcolor=blue]{hyperref}
\usepackage{fontenc}
\usepackage{upgreek}
\usepackage[utf8]{inputenc}
\usepackage{mathrsfs}
\usepackage{booktabs}
\usepackage{pifont}
\usepackage{amsmath}
\usepackage{multirow}
\usepackage{bm}
\usepackage{wasysym}
\usepackage{caption}
\usepackage{subcaption}
\usepackage{euscript}
\usepackage{natbib}
\usepackage{lipsum}
\usepackage{xcolor}

\usepackage{orcidlink}

\usepackage{subcaption}
\usepackage{ragged2e}
\DeclareCaptionJustification{justified}{\justifying}
\newcommand{\jfm}[1]{\textcolor{blue}{#1}}

\begin{document}

\title{Effect of directionality on extreme wave formation during nonlinear shoaling}

\author{Jie Zhang \orcidlink{0000-0003-0794-2335}}
\affiliation{College of Shipbuilding Engineering, Harbin Engineering University, Harbin 150001, China}

\author{Yuxiang Ma \orcidlink{0000-0003-4314-0428}}
\affiliation{State Key Laboratory of Coastal and Offshore Engineering, Dalian University of Technology, Dalian 116023, PR China}

\author{Jiawen Sun}
\affiliation{National Marine Environmental Monitoring Center, Dalian 116023, China}
   
\author{Limin Huang \orcidlink{0000-0002-7944-2754}}
\email{huanglimin@hrbeu.edu.cn}
\affiliation{College of Shipbuilding Engineering, Harbin Engineering University, Harbin 150001, China}

\author{Michel Benoit \orcidlink{0000-0003-4195-2983}}
\affiliation{EDF R\&D, Laboratoire National d’Hydraulique et Environnement (LNHE), Chatou 78400, France}
\affiliation{LHSV, ENPC, Institut Polytechnique de Paris, EDF R\&D, Chatou 78400, France}

\author{Saulo Mendes \orcidlink{0000-0003-2395-781X}}
\affiliation{School of Civil and Environmental Engineering, Nanyang Technological University, Singapore}

\begin{abstract}
Recent studies have shown that, in coastal waters where water depth decreases significantly due to rapid bathymetric changes, the non-equilibrium dynamics (NED) substantially increases the occurrence probability of extreme (rogue) waves. Nevertheless, research on depth-induced NED has been predominantly confined to unidirectional irregular waves, while the role of directionality remains largely unexplored. The scarce studies on multidirectional waves mainly rely on numerical simulations and have yielded conflicting results. In this work, we report on an experimental investigation of wave directionality on the depth-induced non-equilibrium wave statistics. 
High-order statistical moments, skewness and kurtosis, are used as proxies for the non-equilibrium wave response. Our results indicate that the directional spreading has a minor effect on decreasing the maximum values of these statistical moments. In contrast, the incidence direction plays a significant role in the non-equilibrium wave response, which is attributed to the effective bottom slope. 
\end{abstract}


\maketitle

\section{Introduction}\label{sec:intro}

Rogue waves, often associated with severe disasters and casualties, were once dismissed as maritime folklore until numerous encounters with vessels and offshore platforms were documented \citep{Draper1964, Didenkulova2020, Didenkulova2023}. These anomalous ocean waves (with wave height 2 or 2.2 times the significant wave height, by definition) occur with unexpectedly high probability \citep{Dysthe2008, Mori2024, Bitner2024}. 

Rogue wave formation essentially results from wave energy focusing through linear or nonlinear mechanisms. Numerous hypotheses regarding these focusing processes have been proposed \citep{Kharif2009, Onorato2013, Adcock2014, Dematteis2019}, which can be systematically categorized based on the role of nonlinearity. Linear focusing mechanisms include dispersive focusing and spatial focusing due to bathymetric or current-induced reflection and refraction \citep{Chien2002}. Nonlinear focusing mechanisms encompass bound wave nonlinearity \citep{Fedele2016, Knobler2025}, modulation (Benjamin-Feir) instability \citep{Benjamin1967a, Onorato2009, He2022, Zhai2025}, and non-equilibrium dynamics (NED) due to significant and rapid environmental changes \citep{Trulsen2018}.

Recently, NED has attracted considerable interest as it provides a universal explanation of rogue wave formation under rapidly varying wind, current, or bathymetric conditions \citep{Trulsen2012, Annenkov2015, Toffoli2017, Zhang2023, Mendes2025}, or under unrealistic initial conditions \citep{Shemer2010, Tang2022}. In particular, as being relevant in coastal areas, NED induced by drastic depth reduction has been intensively investigated in both numerical \citep{Zeng2012, Gramstad2013, Kashima2014, Viotti2014, Zheng2020, Lawrence2021b, Zhang2024c, Zhang2024, Zhang2024b} and experimental flumes \citep{Trulsen2012, Trulsen2020, Bolles2019, Zhang2019, Wang2020, Lawrence2021a, Li2021b, Samseth2025}, focusing on statistical distributions of free surface elevation and kinematics, evolution of statistical moments, and wave forces on structures. Theoretical investigations on NED have been conducted from both stochastic \citep{Majda2019,Majda2020,Mendes2022,Mendes2023} and deterministic perspectives \citep{Li2021, Moss2025}.

Research on NED in three-dimensional (3D) cases is gaining significant momentum. Using fully nonlinear potential flow models, \citet{Lawrence2022a} and \citet{Draycott2025} investigated the interplay between long-crested waves and two-dimensional bathymetry (circular shoals), both of which show a significant kurtosis enhancement due to NED. With the high-order spectral method, \citet{Ducrozet2017} showed that directional spreading strongly suppresses rogue wave formation due to depth variation, while \citet{Tang2023} observed a weaker directional effect. Conversely, \citet{Lyu2023} showed with a 2D depth-modified nonlinear Schr{\"o}dinger equation that short-crested wave fields with a larger spreading angle may increase the occurrence probability of extreme waves. Remarkably, \citet{Mei2023} found through a fully nonlinear Boussinesq model that directional spread can decrease the excess kurtosis in intermediate water but also increase it in shallow water. Evidently, previous studies on NED of 3D waves are either based on numerical simulations or have been constrained by experimental facility limitations \citep{Tang2023}. Contradictory conclusions indicate that the effect of the wave directionality on rogue wave formation remains unclear. To address this knowledge gap, we conduct a systematic, large-scale experimental study of NED in wave fields with varying directional spreading and incident direction. We explore how the non-equilibrium parameters, skewness and kurtosis, evolve after a rapid depth variation.

\begin{figure*}
\centering
    \includegraphics[width=0.85\textwidth]{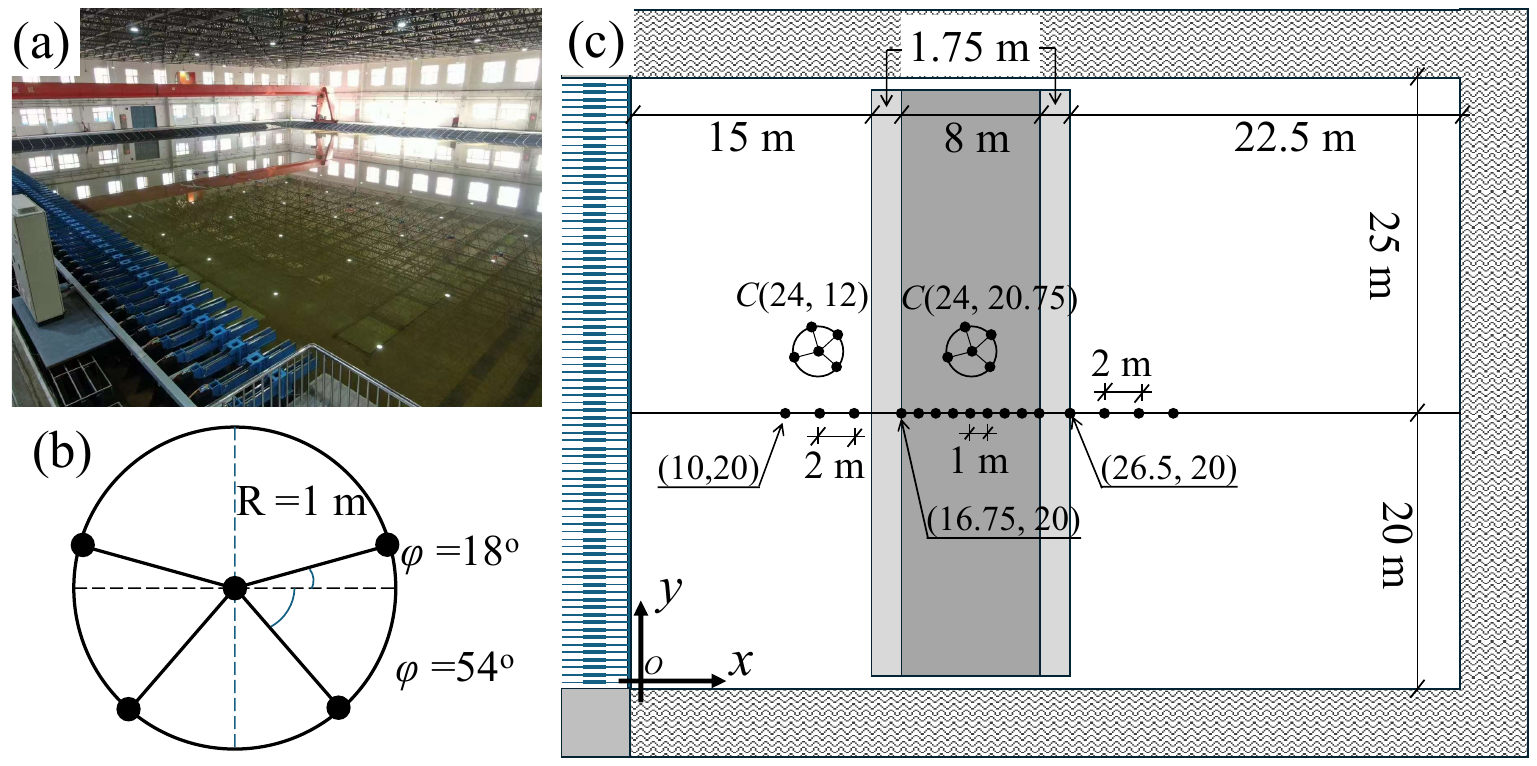}
    \caption{NMEMC wave tank (a); The layout of wave gauge array following \citet{Nwogu1989} for the estimation of directional spectrum (b); The experimental wave tank and locations of the wave gauges (c).}
    \label{fig:bathy}
\end{figure*}

\section{Experimental Setup}\label{sec:theory} 

The experimental campaign was conducted at the Multi-functional Test Basin of the National Marine Environmental Monitoring Center (NMEMC) in Dalian, China. The wave tank is 49~m long, 47~m wide, and 1.2~m deep. Waves are generated by a snake-type (consisting of 80$\times$0.5~m paddles) multi-directional wave maker installed on one shorter side of the tank. Porous media damping zones along the opposing end and lateral boundaries are set to minimize wave reflection.

In the wave tank, an isosceles trapezoidal prism parallel to the wavemaker was installed on an otherwise flat bottom. It consists of a central flat section ($8$~m in width and $0.36$~m in height) flanked by two transitional slopes (with a horizontal extent of $1.75$~m, thus a gradient of $1/4.86$). The upslope of the trapezoidal prism starts $15$~m away from the wavemaker. During the experimental campaign, the water depth near the wavemaker was set to $h_1=0.61$~m, thus $h_2=0.25$~m over the submerged structure. Twenty-six resistance-type wave gauges were deployed during the tests, with a sampling frequency $f_s=50$~Hz. Two arrays of five probes with the same configuration as introduced in \citet{Nwogu1989} were used to estimate the directional spectra before and above the submerged bar. The remaining probes are adopted to capture the wave evolution. The schematic of the experimental basin, as well as the locations of the wave gauges, are shown in Fig.~\ref{fig:bathy}. 

\begin{figure*}
\centering
    \includegraphics[width=0.999\textwidth]{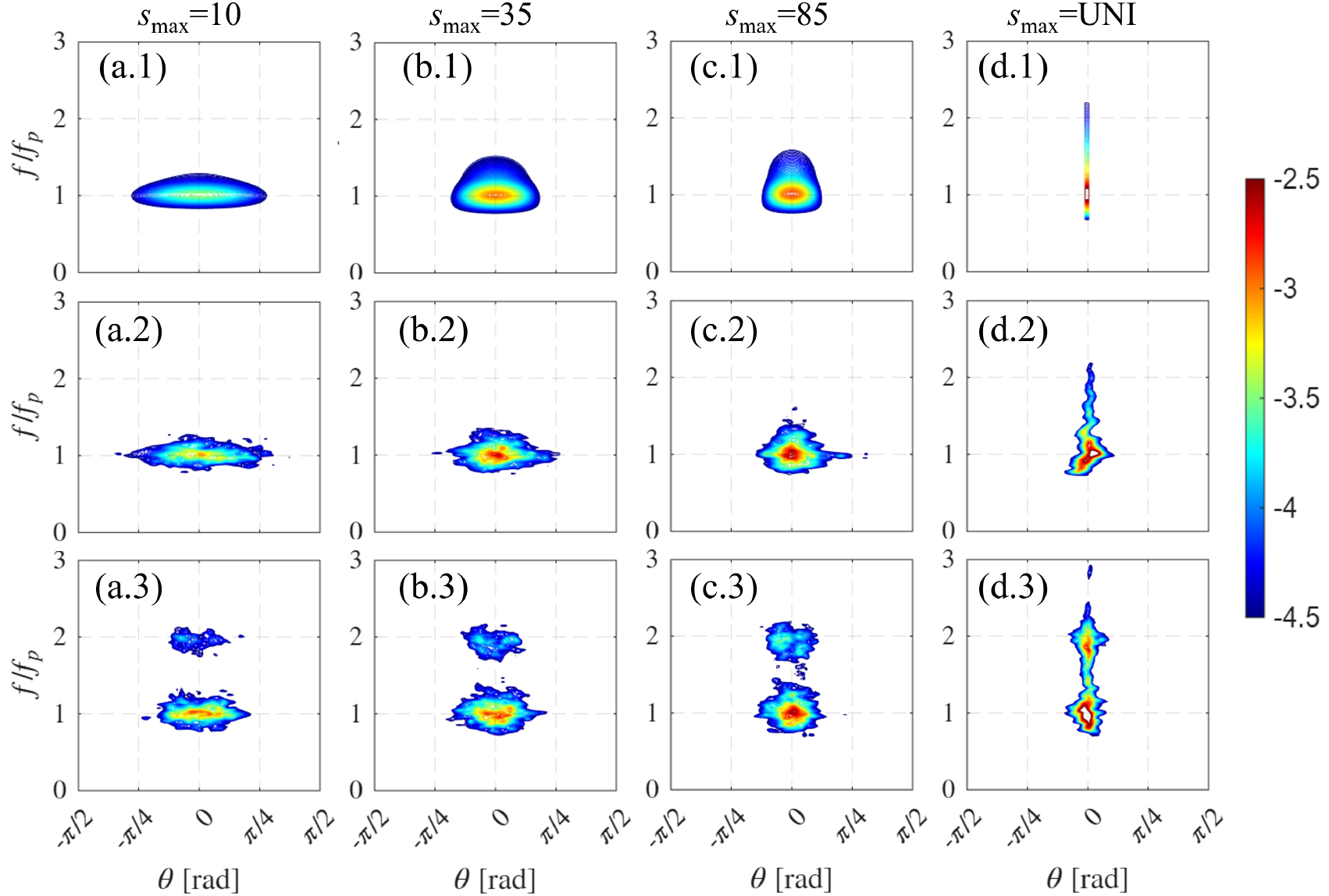}
    \caption{Target directional spectra for the normal incident cases with $s_{\textrm{max}}=10,\, 35,\, 85$ and UNI in panels (a.1--b.1); the corresponding experimental spectra measured offshore (a.2--d.2) and atop of the bar (a.3--d.3).}
    \label{fig:3Dspec}
\end{figure*}
The wave fields are described by their directional spectrum $S(f,\theta)=S_J(f)D(\theta|f)$, with $S_J(f)$ denoting the JONSWAP spectrum and $D(\theta|f)$ the directional spreading function: 
\begin{equation}
\label{eq:jonswap}
  S_J(f)=\frac{\alpha_J{g}^2}{\left(2\pi\right)^4}\frac{1}{f^5}\exp{\left[-\frac{5}{4}\left(\frac{f_p}{f}\right)^4\right]}\gamma^{\exp{\left[-\frac{\left(f-f_p\right)^2}{2\left(\sigma_J f_p\right)^2}\right]}},
\end{equation}
where $f_p=1/T_p$ denotes the spectral peak frequency, $\alpha_J$ the spectral energy parameter, $\gamma$ the peakedness parameter, and $\sigma_J$ width parameter, $\sigma_J=0.07$ for $f<f_p$ and $\sigma_J=0.09$ for $f\geq f_p$. Here, the classical Mitsuyasu-type directional spreading function \citep{Goda2010} is adopted:
\begin{equation}
D(\theta|f) = \frac{2^{2s-1}}{\pi}\frac{\Gamma^2(s+1)}{\Gamma(2s+1)}\cos^{2s}\left(\frac{\theta-\theta_{\textrm{inc}}}{2}\right),
\end{equation}
with $\Gamma$ denoting the Gamma function, $\theta_{\textrm{inc}}$ the dominant wave direction, and $s$ the frequency-related angular spreading parameter, $s(f)=s_{\max} (f/f_p)^{5}$ for $f \leq f_p$ and $s=s_{\max} (f/f_p)^{-2.5}$ for $f\geq f_p$. The peak value of $s$, $s_{\max}$, governs the width of the directional spreading. To avoid phase locking and ensure ergodicity of the generated wave field, we adopt the modified double summation approach for wave generation, which associates a random phase $\phi_{ij}\in[0,\, 2\pi]$ to each frequency-direction bin $(f_{ij},\theta_j)$ \citep{Luo2020}:
\begin{equation}
    \eta(x,y,t)  = \sum_{i=2}^{N_f} \sum_{j=1}^{N_{\theta}} a_{ij}\cos\left(k_{ij}x\cos\theta_j - 2\pi f_{ij}t + \phi_{ij}\right),
\end{equation}
where
\begin{align}
    \begin{cases}
    f_{ij} = \hat f_i - \frac{1}{2}\Delta  f + {\left(j-1+{R}_{ij}\right)}\Delta  f/N_{\theta}, \\
    \hat f_i=( f_{i-1}+ f_i)/2,
\end{cases}
\end{align}
with $N_f$, $N_{\theta}$ denoting the discretization points in frequency and direction, respectively, and $\Delta f$, $\Delta \theta$ being the uniform intervals. The cut-off range for the directional spectrum $S(f,\theta)$ is $f_i\in[0.5f_p,\, 3.5f_p]$ and $\theta_j\in[-\pi,\,\pi]$. $a_{ij}$, $k_{ij}$, and $\phi_{ij}$ denote the amplitude, wavenumber, and random phase of the component harmonics, respectively. $R_{ij}$ denotes a random number in $[0, 1]$ with a uniform probability distribution.

The configurations of incident wave fields are listed in table~\ref{tab:cases}, together with non-dimensional parameters, wave steepness $\varepsilon \equiv \sqrt{2}k_p\sigma$ and relative water depth $\mu \equiv k_p h$, where $\sigma$ denotes the standard deviation of the measured free surface elevation (FSE) and $k_p$ the spectral peak wavenumber. The subscripts $0$ and $f$ denote deeper-water and shallower-water quantities, respectively. In all cases, the waves generated offshore are of relatively mild nonlinearity.
Three peak periods $T_p=[1.2,\, 1.6,\, 2.0]$~s were examined, corresponding to weak, intermediate, and strong non-equilibrium scenarios, respectively. Here, the cases with $T_p=1.6$~s and 2.0~s are shown. The peakedness parameter $\gamma=3.3$ was maintained constant across all cases. Three directional spreading conditions $s_{\max}=[10, \, 35, \, 85]$ were tested, in which 95\% spectral energy concentrated within the range $[-0.39, 0.39]\pi$, $[-0.22, 0.22]\pi$, and $[-0.14, 0.14]\pi$, respectively. The B2 and B4 cases were tested with three oblique incidence conditions, $\theta_{\textrm{inc}}=\pi/12$, $\theta_{\textrm{inc}}=\pi/6$, and $\theta_{\textrm{inc}}=\pi/4$. Note that the wave gauges are shifted from $y=20$~m to $y=24$~m in the cases with $\theta_{\textrm{inc}}=\pi/4$ to capture wave evolution. Each experimental run lasts 6 minutes for wave generation and data acquisition, allowing for the limitation of energy accumulation due to long-wave reflection. To ensure statistical stability, several runs with different random phase seeds were performed such that the total sample duration exceeded 5,000$T_p$ for each case.
\begin{table*}
\begin{center}
\begin{tabular}{cccccccccc}
\cline{1-10}
\multirow{2}{*}{Case} & \multirow{2}{*}{$T_p$ [s]} & \multirow{2}{*}{$s_\textrm{max}$} & \multirow{2}{*}{$\theta_{\textrm{inc}}$ [rad]} & 
\multicolumn{3}{c}{Deeper zone ($h_1=0.61$~m)}  & \multicolumn{3}{c}{Shallower zone ($h_2=0.25$~m)} \\[0.5ex]
 & & & & {$H_{s,0}$ [m]} & $\varepsilon_0$ & {$\mu_0$} &  {$H_{s,f}$ [m]} & $\varepsilon_f$ & {$\mu_f$} \\[0.5ex]
\cline{1-10} 
A1 & \multirow{4}{*}{1.6} & 10 & \multirow{4}{*}{$0$} & 0.032 & 0.021 & \multirow{4}{*}{1.165}  & 0.033 & 0.031 & \multirow{4}{*}{0.671} \\
A2 & & 35  &  &  0.034 & 0.022 &  & 0.036 & 0.034 &  \\
A3 & & 85  &  &  0.034 & 0.023 &  & 0.038 & 0.035 &  \\
A4 & & UNI &  &  0.034 & 0.023 &  & 0.038 & 0.036 &  \\
\cline{1-10}
B1 & \multirow{4}{*}{2.0} & 10 & 0 & 0.044 & 0.022 & \multirow{4}{*}{0.873}  & 0.050 & 0.037 & \multirow{4}{*}{0.524}   \\
B2 & & 35  & $[0, \pi/12, \pi/6, \pi/4]$ & 0.046 & 0.023 &   & 0.053 & 0.039 &  \\
B3 & & 85  & 0 & 0.046 & 0.023 &   & 0.054 & 0.040 &  \\
B4 & & UNI & $[0, \pi/12, \pi/6, \pi/4]$ & 0.048 & 0.024 &   & 0.055 & 0.039 &  \\[0.5ex]
\cline{1-10}
\end{tabular}
\caption{Incident wave conditions and key non-dimensional parameters \label{tab:cases}}
\end{center}
\end{table*}
\begin{figure*}
\centering
    \includegraphics[width=0.999\textwidth]{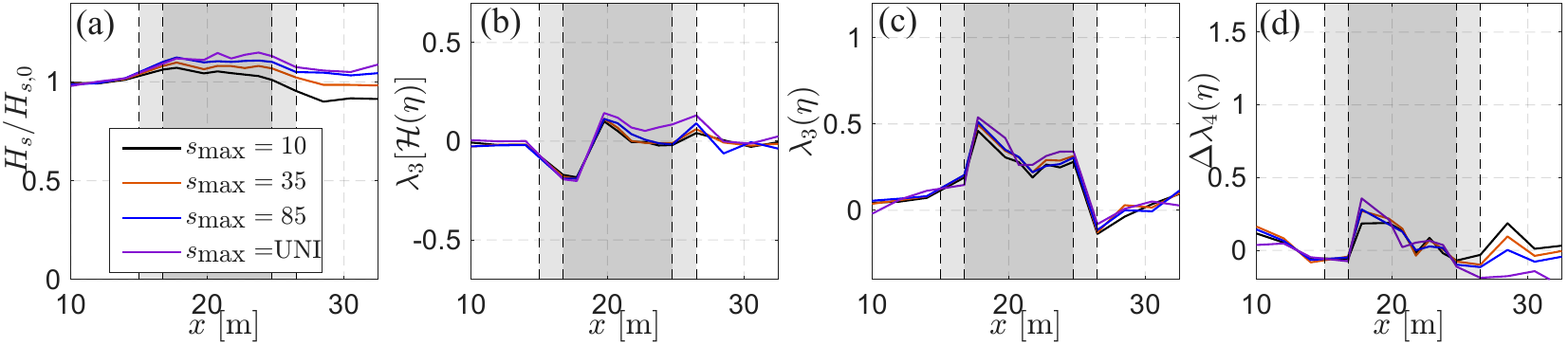}
    \caption{Spatial evolution of normalized significant wave height (a), asymmetry parameter (b), skewness (c), and net change of kurtosis (d) of Cases A1--A4 (all with normal incidence, $\theta_{\textrm{inc}}=0$).}
    \label{fig:Smax_CasA}
\end{figure*}
\begin{figure*}
\centering
    \includegraphics[width=0.999\textwidth]{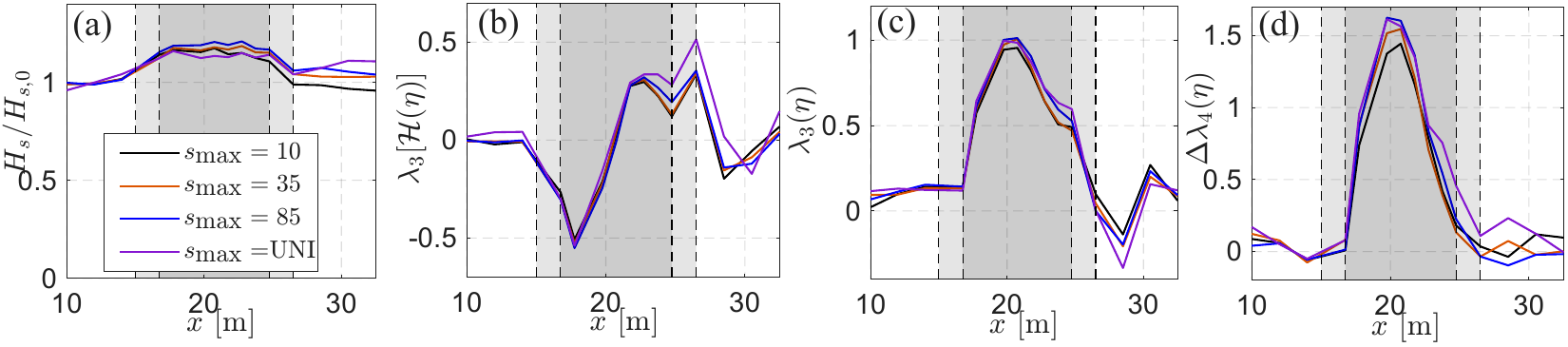}
    \caption{Same as \jfm{Fig.}~\ref{fig:Smax_CasA}, but for normal incident directional wave cases ($\theta_{\textrm{inc}}=0$) of Cases B1--B4.}
    \label{fig:Smax_CasB}
\end{figure*}
\begin{figure*}
\centering
    \includegraphics[width=0.999\textwidth]{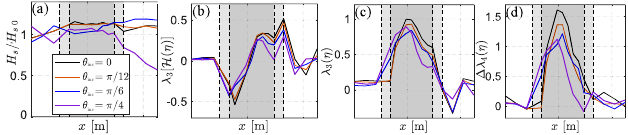}
    \caption{Same as \jfm{Fig.}~\ref{fig:Smax_CasA}, but for oblique incident unidirectional cases ($s_\textrm{max}=$UNI).}
    \label{fig:theta0_UNI}
\end{figure*}
\begin{figure*}
\centering
    \includegraphics[width=0.999\textwidth]{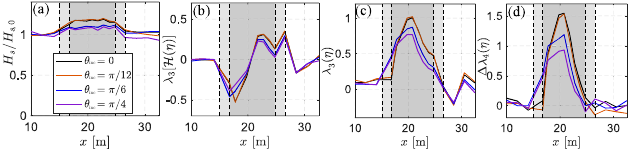}
    \caption{Same as \jfm{Fig.}~\ref{fig:Smax_CasA}, but for oblique incident directional cases ($s_\textrm{max}=35$).}
    \label{fig:theta0_S35}
\end{figure*}
\begin{figure*}
\centering
    \includegraphics[width=0.9\textwidth]{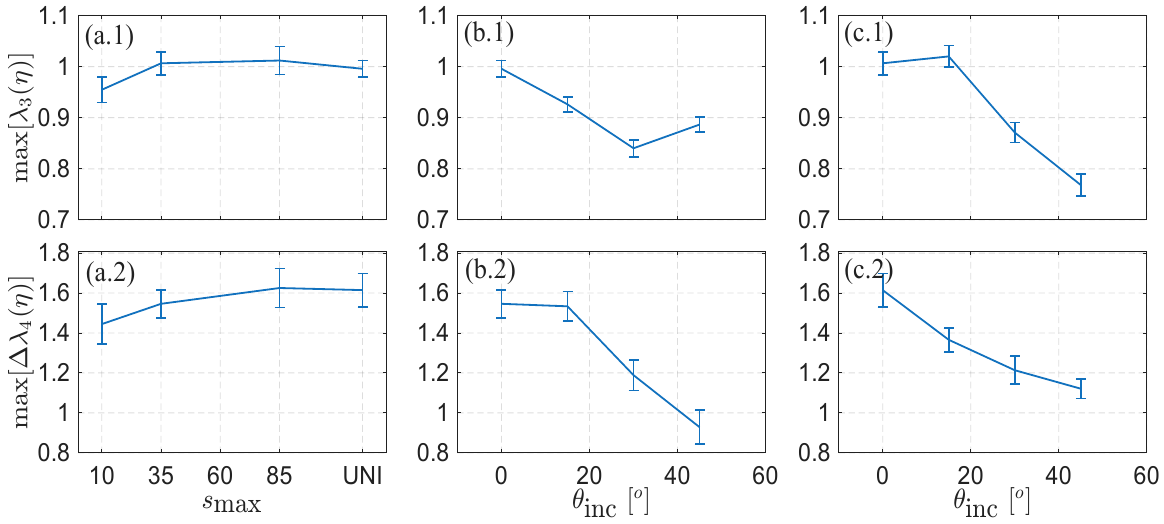}
    \caption{Maximum values of $\lambda_3$ and $\Delta\lambda_4$ as functions of $\theta_{\textrm{inc}}$ and $s_{\textrm{max}}$, in normal incident directional wave cases (a), oblique incident unidirectional wave cases (b), and oblique incident directional wave cases with $s_\textrm{max}=35$ (c).}
    \label{fig:SK_all_case}
\end{figure*}

\section{Results and Discussions \label{ssec:results}} 
To verify the quality of directional wave field generation in the offshore area and to show the spectral evolution as waves propagate over the bar, two wave gauge arrays were deployed in the wave tank. Fig.~\ref{fig:3Dspec}(a.1--d.1) shows the target spectra of the normal incidence B1--B4 cases with $\theta_{\textrm{inc}}=0$ and four values of $s_{\textrm{max}}$, the corresponding spectra measured offshore (a.2--d.2), and atop of the bar (a.3--d.3). The directional estimation is achieved by using a classical iterative maximum likelihood approach. In Fig.~\ref{fig:3Dspec}, the measured and target spectra are in good agreement offshore, indicating that the directional wave fields were generated properly. At the top of the bar, the second-order harmonics are excited significantly after shoaling, consistent with the observations in long-crested wave scenarios. Moreover, it is noticed that the enhancement of the second-order harmonics slightly decreases as $s_{\textrm{max}}$ decreases. The same good wavemaking quality also applies to other cases with normal/oblique incidence, which are not displayed due to the length limit of Rapids papers. 

Following previous works \citep[see][for example]{Onorato2005, Trulsen2020, Li2021}, we focus on the evolution of third and fourth-order statistical moments, skewness and kurtosis, of the FSE, which characterize the non-Gaussianity of sea states. Skewness, $\lambda_3(\eta)\equiv \langle (\eta-\langle\eta\rangle)^3\rangle/\sigma^3$ with $\langle\cdot\rangle$ being an averaging operator, measures the asymmetry of the wave profile in the vertical direction. Kurtosis, $\lambda_4(\eta)\equiv \langle (\eta-\langle\eta\rangle)^4\rangle/\sigma^4$, serves as a proxy of rogue wave intensity. In addition, the asymmetry parameter, $\lambda_3[\mathcal{H}(\eta)]$, measures the horizontal asymmetry of the wave profile, with $\mathcal{H}(\cdot)$ denoting Hilbert transform. In the following, the spatial evolutions of the statistical parameters, including the normalized significant wave height, asymmetry parameter, skewness, and the net change of kurtosis $\Delta\lambda_4$ (with the offshore mean kurtosis subtracted), are discussed. 

Fig.~\ref{fig:Smax_CasA} shows the spatial evolution of the statistical parameters in cases A1--A4 with different directional spreadings. Fig.~\ref{fig:Smax_CasA}(a) shows that the normalized significant wave height is slightly increased over the bar due to shoaling. Fig.~\ref{fig:Smax_CasA}(b) and (c) indicate that the NED effects result in increased asymmetry of the mean wave profile in both horizontal and vertical directions. In Fig.~\ref{fig:Smax_CasA}(d), kurtosis is moderately enhanced over the bar, implying higher rogue wave probability. These observations are in line with those reported in unidirectional wave studies \citep[see][for instance]{Lawrence2021a, Zhang2019, Zheng2020}. Fig.~\ref{fig:Smax_CasB} shows the statistical parameter evolution in the normal incident B1--B4 cases with varying $s_{\textrm{max}}$. The evolution trend is very similar to that of cases A1--A4, except for a more pronounced NED response, i.e., higher peaks of skewness and kurtosis due to higher steepness atop the shoal $\varepsilon_f$ and smaller relative water depth $\mu_f$ (see table \ref{tab:cases}). 

Combining Figs.~\ref{fig:Smax_CasA} and ~\ref{fig:Smax_CasB}, we conjecture that directional spreading induces only minor changes in the statistical parameters, especially on the top of the bar, where waves are out of equilibrium. 
The normalized significant wave height ($H_s / H_{s,0}$) is slightly disturbed by the directional spread in the shoaling zone, while strongly disturbed in the de-shoaling zone, which is presumably due to wave reflection and refraction. The skewness and the asymmetry parameter are almost unchanged for different $s_{\textrm{max}}$ because of the similar level of wave nonlinearity (steepness) over the bar. Only minor differences are seen over the down-slope area, and in the deeper flat region after it. Meanwhile, the net change of kurtosis is mildly reduced for broader directional spreading. In B4 case with $s_{\textrm{max}}=\textrm{UNI}$ and $\theta_{\textrm{inc}}=0$, the total kurtosis maximum value achieves 4.61 over the bar, while in B1 case with $s_{\textrm{max}}=\textrm{UNI}$ and $\theta_{\textrm{inc}}=0$, a comparable value, 4.44 is achieved. Both indicate a strongly non-Gaussian sea state resulting from non-equilibrium wave evolution, and a much higher probability of rogue waves than that expected in a Gaussian sea state.

Fig.~\ref{fig:theta0_UNI}(a)--(d) show the evolution of statistical parameters for the oblique incident unidirectional wave B4 cases, with $s_{\textrm{max}}=\textrm{UNI}$ and varying $\theta_{\textrm{inc}}$. It is worth mentioning that in panel (a), the normalized significant wave height decreases dramatically after the down-slope for the case with $\theta_{\textrm{inc}}=\pi/4$. This is due to a limitation of the experimental facility: The wavemaking paddles are installed on one side of the wave tank, thus leaving the last four wave gauges outside the effective experimental zone. Despite the decrease in significant wave height, this case is included for discussion anyway, as the non-Gaussian behaviour atop the bar is of more interest and is not affected by the facility's limitation. From Fig.~\ref{fig:theta0_UNI}(c)--(d), we notice that, not only the maximum values, but also the locations where the maxima of skewness and kurtosis are achieved, vary considerably with $\theta_{\textrm{inc}}$. We attribute this to the effective slope gradient, which is crucial for the wave non-equilibrium evolution. The effective slope, defined as the ratio of the bar height and horizontal upslope length in the direction of wave propagation, reads $1/4.86$, $1/5.03$, $1/5.61$, and $1/6.87$ for $\theta_{\textrm{inc}}=0$, $\pi/12$, $\pi/6$, and $\pi/4$, respectively.

Fig.~\ref{fig:theta0_S35}(a--d) show the evolution of statistical parameters in cases that comprise both directionality and oblique incidence. In this case, as the wave fields are of a relatively broad band in directional spreading, wave energy can be transmitted to a broader range in the wave tank, the significant wave height in the case with $\theta_{\textrm{inc}}=\pi/4$ does not reduce as significantly as in Fig.~\ref{fig:theta0_UNI}(a).
The evolution trends of asymmetry parameter, skewness, and kurtosis in Fig.~\ref{fig:theta0_S35}(b--d) are very similar to those observed in Fig.~\ref{fig:theta0_UNI}(b--d), their maximum values increase with increasing effective slopes, and the locations where the maxima are achieved shift towards downstream. Inversely, the out-of-equilibrium sea states adapt faster to the new water depth when the effective slope is milder.

To further illustrate the role played by the directionality and the oblique incidence on the non-equilibrium wave dynamics, in Fig.~\ref{fig:SK_all_case}, the maximum values of skewness and kurtosis in B1--B4 cases are extracted and plotted as functions of $s_{\textrm{max}}$ and $\theta_{\textrm{inc}}$. It is clearly shown in Fig.~\ref{fig:SK_all_case}(a.1) and (a.2) that increasing the directional spreading $s_{\textrm{max}}$ results in only about 10\% lower skewness and kurtosis than in the unidirectional case. Our result is in line with the observations of \citet{Tang2023}, and in contrast to those of \citet{Lyu2023}. The main reason is that in the latter work, the relative water depth of the shallower region is around $\mu_f = 1.1$. Although it falls within the typical NED range $0.5 \leqslant \mu_f \leqslant 1.3$ \citep{Trulsen2020, Zhang2023}, it corresponds to weakly non-equilibrium wave evolution. 
Thus, the effect of directionality on the non-equilibrium response (e.g., the skewness and net change in kurtosis) is weak and comparable to the statistical error atop the bar, and stronger downslope. Moreover, the nonlinear Schr{\"o}dinger equation is limited in describing high-order bound harmonics. In Fig.~\ref{fig:SK_all_case}(b) and (c), the effects of oblique incidence, which have been rarely discussed, are shown. We find that the incident angle plays an important role in the magnitude of the non-equilibrium wave response. This is attributed to the effective bottom slope. However, it should be noted that the wave gauge locations were chosen to capture the maximum skewness and kurtosis for the normal incident case, but are likely not optimal for cases with oblique incidence. Therefore, the maximum values of skewness and kurtosis could be underestimated to some extent. Whereas, from the relatively smooth evolution of skewness in Figs.~\ref{fig:theta0_UNI}(c) and \ref{fig:theta0_S35}(c), we believe that the conclusion will not be overturned when the ``actual'' maximum values of skewness and kurtosis are considered. Further investigations of oblique incident cases with more wave gauges over the bar, using numerical simulations and statistical distributions, will be discussed in a subsequent paper.

\section{Conclusions \label{sec:conclusion}}

The present study provides a comprehensive experimental investigation into the effect of wave directionality on extreme wave formation during nonlinear shoaling, focusing on non-equilibrium dynamics (NED) induced by rapid depth changes. Unlike previous research, which predominantly explored unidirectional irregular waves or directional waves with small directional spread, this work systematically examines multidirectional wave fields using a large-scale wave tank over realistic steep smooth slopes. The oblique incident angle $\theta_{\textrm{inc}}$ was chosen up to $\pi/4$ (the limit of our experimental setup), and a wide range of directional spreading with $s_{\textrm{max}}=10$ to unidirectional condition has been tested. Our results indicate that directional spreading has a minor impact on reducing statistical moments such as skewness and kurtosis for a relatively steep slope, contrasting with previous numerical studies that suggest evident suppression of rogue wave formation due to energy dispersion with directionality.

In contrast to the literature's limited exploration of oblique incidence, the present study highlights the significant role of the incidence direction in NED, driven by the effective bottom slope. This finding extends beyond previous works, which focused on normal incidence, by showing that obliqueness significantly suppresses non-Gaussian behaviour and rogue wave formation. These experimental results, using a Mitsuyasu-type directional spreading function and varying incidence angles, offer a more nuanced understanding than the contradictory numerical simulations, suggesting that future research should prioritize experimental validation and finer resolution studies to clarify these dynamics.

\section{Acknowledgements}
The authors wish to acknowledge Dr. Zhou Guanting from NMEMC and Dr. Tan Ting from DUT for their technical help during the experimental campaign.

\section{Declaration of interests}
The authors report no conflict of interest.

\bibliography{Maintext}

\end{document}